**This article is a preprint of**

Holden, MH., Plagányi, E, Fulton, EA, Campbell, AB, Janes, R, Lovett, RA, Wickens, M, Adams, MP, Botelho, LL, Dichmont, CM, Erm, P, Helmstedt, KJ, Heneghan, RF, Mendiolar, M, Richardson, AJ, Rogers, JGD, Saunders, K, Timms, L. (2024) Cost-benefit analysis of ecosystem modelling to support fisheries management. *Journal of Fish Biology*.

https://doi.org/10.1111/jfb.15741

**Article Type:** *Perspective*

**Title:** Cost-benefit analysis of ecosystem modelling to support fisheries management

Matthew H. Holden[1,2], Eva E. Plagányi[3,4], Elizabeth A. Fulton[4,5], Alexander B. Campbell[6], Rachel Janes[6], Robyn A. Lovett[6], Montana Wickens[6], Matthew P. Adams[7,8,9], Larissa Lubiana Botelho [7,12], Catherine M. Dichmont[10], Philip Erm[11], Kate J Helmstedt[7,12], Ryan F. Heneghan[7], Manuela Mendiolar[1,2], Anthony J. Richardson[1,2,3], Jacob G. D. Rogers[3], Kate Saunders[8,13], Liam Timms[1,2]

1. School of Mathematics and Physics, University of Queensland, St Lucia QLD, 4072, Australia
2. Centre for Biodiversity and Conservation Science, University of Queensland, St Lucia QLD, 4072, Australia
3. CSIRO Environment, Brisbane, Queensland, 4072, Australia
4. Centre for Marine Socioecology, University of Tasmania, Hobart, Australia
5. CSIRO Environment, Hobart, Tasmania, 7001, Australia
6. Fisheries Queensland, Department of Agriculture and Fisheries, Brisbane Qld 4001, Australia
7. School of Mathematical Sciences, Queensland University of Technology, Brisbane QLD, 4001, Australia
8. Centre for Data Science, Queensland University of Technology, Brisbane QLD, 4001, Australia
9. School of Chemical Engineering, The University of Queensland, St Lucia QLD, 4072, Australia

10. Cathy Dichmont Consulting, Banksia Beach, QLD, 4507, Australia
11. Conservation Science Group, Department of Zoology, University of Cambridge, Cambridge CB2, UK
12. Securing Antarctica's Environmental Future, Queensland University of Technology, Brisbane QLD 4001, Australia
13. Department of Econometrics and Business Statistics, Monash University, VIC, Australia

**Corresponding Author:** Matthew H Holden, m.holden1@uq.edu.au

**Abstract:** Mathematical and statistical models underlie many of the world's most important fisheries management decisions. Since the 19th century, difficulty calibrating and fitting such models has been used to justify the selection of simple, stationary, single-species models to aid tactical fisheries management decisions. Whereas these justifications are reasonable, it is imperative that we quantify the value of different levels of model complexity for supporting fisheries management, especially given a changing climate, where old methodologies may no longer perform as well as in the past. Here we argue that cost-benefit analysis is an ideal lens to assess the value of model complexity in fisheries management. While some studies have reported the benefits of model complexity in fisheries, modeling costs are rarely considered. In the absence of cost data in the literature, we report, as a starting point, relative costs of single-species stock assessment and marine ecosystem models from two Australian organizations. We found that costs varied by two orders of magnitude, and that ecosystem model costs increased with model complexity. Using these costs, we walk through a hypothetical example of cost-benefit analysis. The demonstration is intended to catalyze the reporting of modeling costs and benefits.

**Keywords:** ecosystem modeling, stock assessments, fisheries modeling, population dynamics, tactical decision-making

## Introduction

Both simple and complex models provide major societal benefits. They allow us to make more objective, less-biased decisions (McCarthy et al., 2004), which can produce better outcomes than decisions based solely on subjective judgement (Czaika & Selin, 2017; Holden & Ellner, 2016). But despite their clear benefits, when it comes to high-stakes decision-making, both simple and complex models carry serious costs. On the one hand, simple models are often unable to predict unexpected perverse outcomes caused by human action. From the collapse of fisheries to financial markets, the blind use and misapplication of simple models have sometimes backfired (Stewart, 2012; Walters & Maguire, 1996). Conversely, large complex models can be challenging to implement, calibrate, and interpret. While they are capable of foreshadowing unexpected outcomes, their predictions can be inaccurate due to both overfitting limited datasets and their sensitivity and instability in response to small perturbations in parameter values, model structure, and initial conditions. Complex models, therefore, require considerable resources to refine and calibrate before they provide sensible output. Given these drawbacks and potential benefits of model complexity, how complex should models be? In the case of fisheries, climate and other environmental drivers are changing, which means that ignoring non-stationarity and ecosystem interactions could lead to poor management outcomes. Now, more than ever, is the time to revisit this question for fisheries management.

For this paper, we will focus on ecosystem models, which we define as models that represent interactions among ecosystem components and processes (Geary et al., 2020). Ecosystem models are often highly complex due to having many state variables, parameters, and nonlinear feedbacks and interactions. While models used by fishery management agencies are often complicated (such as an age-structured model with many parameters and state variables), they are still typically single-species population models (Skern-Mauritzen et al., 2016) with few sources of nonlinear interactions (Fig. 1ab). Historically, only 2 of 1,250 publicly available marine fish stock assessments from the USA, Australia, and 22 international fishery bodies included dynamic, two-way feedbacks between two or more species (Skern-Mauritzen et al., 2016); and only 24 (2%) incorporated ecosystem components as independent or external variables rather than feedbacks. Some researchers have advocated for the use of ecosystem models to provide strategic context to fisheries and thereby improve resource management (Fletcher, Shaw, Metcalf, & Gaughan, 2010; Fulton et al., 2014). However, the uncertainty around model outputs and the high costs of collecting the necessary data to constrain these models make the tactical use of quantitative, whole ecosystem-based stock-assessments for annual decisions infeasible in most instances.

While both single-species and ecosystem models have been used successfully for considering a mix of management levers to support sustainable fisheries (Costello et al., 2016; Fulton et al., 2014; Hilborn & Ovando, 2014; Plagányi et al., 2014), clear guidance on which type of model to use in what circumstances would be valuable. This is especially the case with climate change undermining the fundamental stationarity assumptions of simple models, and potentially reducing their predictive power (Szuwalski & Hollowed, 2016; Karp et al., 2023). Complex models with interactions may become increasingly relevant as they can include climate inputs reflecting biophysical fisheries drivers (Blamey et al., 2022; Thorson et al., 2019). However, there are often competing pressures from different stakeholders to both simultaneously reduce cost and add complexity. Here we propose that cost-benefit analyses can help guide decisions concerning model complexity for fisheries management. In the absence of cost data in the literature, we report relative costs of single-species and marine ecosystem models from two Australian organizations. Using these costs, we walk through a hypothetical example of cost-benefit analysis. The demonstration is intended to catalyse the reporting of modelling costs and benefits.

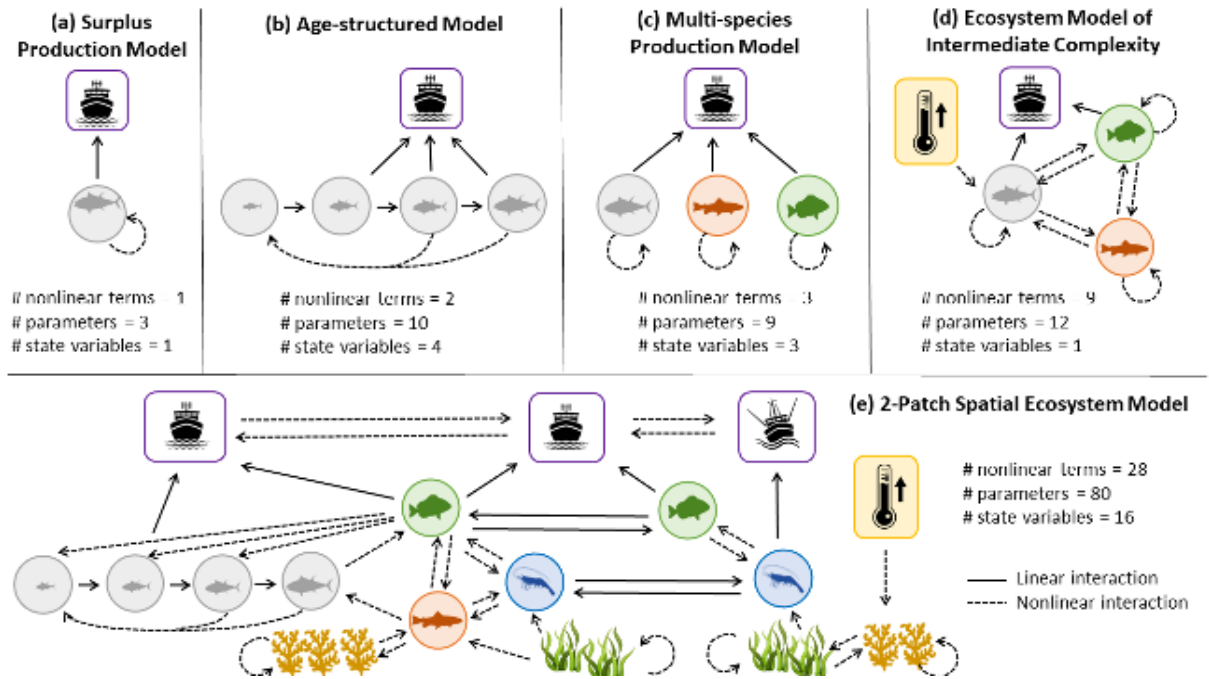

**Fig. 1.** Hypothetical fishery models with an indicative idea of the associated levels of complexity displayed in each diagram along three axes that reflect the number of: (1) state variables; (2) parameters; and (3) nonlinear terms. In these examples, density dependence, demographic structure, interactions among taxonomic groups, biophysical drivers, temporal variability, spatial connectivity, and human behaviour are common mechanisms leading to increased complexity. Of these five models, only the last two meet our definition of ecosystem models, because the third model does not have species interactions. Note that the number of parameters and nonlinear terms for each model type can vary by orders of magnitude depending on the application, the numbers in this figure are only indicative.

**The benefits and costs of model complexity**

The value of model complexity has two parts: (1) benefits achieved from implementing a more complex model; and (2) costs of implementing the added complexity. To justify the use of a complex model over a simple one, benefits must outweigh costs.

**Benefits** of fisheries modelling may be economic (e.g. increased yield, profit, and employment), environmental (e.g. increased biodiversity, or averted greenhouse gas emissions), social (e.g. mental health, and culture), and/or scientific (e.g. research innovation). The use of a more complex model could lead to such benefits through improved accuracy of predictions (Hellweger, 2017) and increased system understanding, leading to better management decisions. For example, increasing complexity in a single species model through demographic structure (Fig 1b) can identify cohort-specific depletion, indicating risks which would otherwise go unnoticed when modelled purely as biomass (Tahvonen, 2008). Similarly, ecosystem models can identify risks of stock degradation through interactions between species and the environment. Such ecosystem models often lead to recommendations that can increase long-run catch when harvested species interact with other species on similar timescales (Burgess et al., 2016), time-varying exogenous variables drive the system (Burgess et al., 2016), or many species contribute substantially to the total catch within a region (Fulton, et al., 2018).

Certain levels of complexity are also required to address specific benefits or answer specific management questions (Hannah et al., 2010). For example, management questions concerning maintaining biodiversity require multi-species models or at least aggregate indicators of diversity (Geary et al., 2020, Fulton et al., 2003). Additionally, in both single-species and ecosystem models, spatial structure is important when there are concerns regarding localised depletion, and/or spatial heterogeneity in biological parameters, harvest trends, or fishing regulations (Punt, 2019). Such models may be required to make decisions at fine spatial scales, or to see different decisions or management levers used in different locations, to increase sustainable, long-term catch over the whole fishery.

It is important to note that adding model complexity can also decrease benefits in fisheries (Puy et al., 2022). For example, if there is insufficient fundamental understanding of the dynamic processes underpinning the structure of a complex model, insufficient data to match the model's complexity, or if the model is poorly constructed, models could produce inaccurate predictions or be sensitive to moderate structural or data changes. Alternatively, simple models can also result in poor management outcomes, resulting in decreased benefits, such as lost catch and revenue over time (Dichmont et al., 2017). Note that there are usually low benefits of doing no modelling, due to foregone catch or increased risk of stock degradation and ecosystem collapse (Dichmont et al., 2017; Holden & Ellner, 2016).

**Costs** of any model include the cost of building, refining, fitting, calibrating, running, and analysing the model, and interpreting model outputs. There are also less obvious costs, such as for acquiring the underlying knowledge to build the model, collecting and processing data for parameterizing and fitting, and communicating complex model results to stakeholders. Typically, we would expect more complex models to be more costly in all these aspects, due to the increased time and more specialized skills required. Complex models may also have higher ongoing future costs, including to maintain the computer code, and collect and analyse more data to validate previous outputs. For this paper we consider opportunity cost, such as unrealized catch (Dichmont et al., 2017), as a negative or reduced benefit, since some negative benefits may not be easily expressed in monetary units. It is also noteworthy that the costs of developing complex models are decreasing due to coordinated global efforts to share code and platforms, as well as new modular stock assessment frameworks under development that will also support incorporation of ecosystem and climate effects (see e.g Punt et al., 2020).

A key impediment to deciding between ecosystem and single species models is the unknown costs of unfamiliar approaches. Knowledge of modelling costs are often hidden in confidential contracts between fisheries organizations and their clients and stakeholders. Therefore, we report cost ranges for implementing three types of models, each with differing complexity, at two Australian fisheries science organizations, Fisheries Queensland and the Commonwealth Scientific and Industrial Research Organization (CSIRO). The model types considered include: single-species stock assessment models (Fig. 1 ab), multi-species *Models of Intermediate Complexity for Ecosystem* assessments (MICE, e.g. Fig. 1d), and complex ecosystem models (Atlantis, e.g. Fig. 1e). MICE typically have three to ten interacting components that capture the essential aspects of the system and are formally fitted to data (Plagányi et al., 2014; Thorson et al., 2019). Atlantis is a detailed, complex, ecosystem model, typically with hundreds of components, including the option to model spatial, temporal, environmental, ecological, biophysical and/or socio-economic processes and interactions (Fulton et al., 2011). In the cost ranges that we report for each type of model, we included the costs of labour associated with data processing, model construction and analysis, and reporting. Other costs, such as, data collection, data entry, or ongoing costs are not included. This is important to note, as it is often difficult to separate modelling from data collection costs because each type of model chosen (or whether a stock assessment will be done) can depend on the available data.

Across the two organisations, single-species stock assessment modelling costs varied by an order of magnitude between projects (0.24 – 2.66 times the median stock assessment cost, Fig. 2). Atlantis was between 6.3 - 42 times the cost of implementing a typical single-species stock assessment model (for a first implementation to a specific fishery, Fig. 2). MICE were 1.6 - 11.6 times as expensive as a typical stock assessment model, but the cheapest MICE, were cheaper than the most expensive single-species stock assessment model (see bar overlap in Fig. 2).

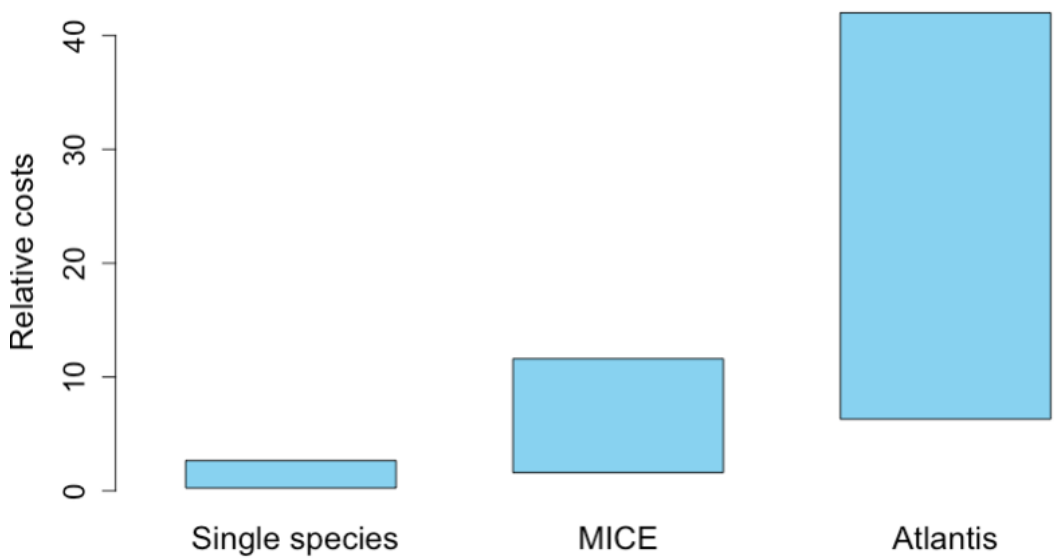

**Fig. 2**: Modelling cost ranges from across Fisheries Queensland (FQ) stock assessments and CSIRO's fisheries modelling contracts, relative to median stock assessment modelling costs. Single species models include surplus production models all the way through to more complex, spatial, age-structured models. Ecosystem models include Models of Intermediate Complexity for Ecosystem assessment (MICE), and one of the world's most complex ecosystem modelling frameworks (Atlantis). Cost is for modeler labour associated with data processing, model construction, analysis, and reporting.

Importantly, the above cost ranges include model development costs, which are especially substantial for complex bespoke models. Ongoing costs of both single-species and ecosystem models can also be substantial due to the time required to process and fit the existing models to new data. However, if the purpose of a model is data-free scenario analysis, there may be much lower ongoing costs to rerun these models without incurring the development cost again. In both cases, the cost of future modelling can decrease due to historical investment and expertise gained.

**Assessing potential benefits relative to costs**

Given a set of benefits and costs, the value of model complexity can be defined in several ways. It could be an explicit function of the benefits and costs, such as benefit minus cost or benefit divided by cost. The goal would then be to choose the model that maximizes this function. If benefits can be easily translated into economic currency (e.g. catch), then benefit minus cost would be appropriate, as it represents the total economic gain from the model. However, for benefits not easily translated into money, the ratio may be more appropriate, and units would be benefit per dollar spent. For example, expected number of at-risk species recovered or maintained above a predefined threshold per dollar spent is interpretable without converting species recovered into dollars. This could allow fisheries scientists and decision-makers to quantify the value of any combination of performance indices such as biodiversity, resilience to stock collapse, or socio-economic benefits.

Ranking *management actions* based on functions of costs and benefits is a common way of informing environmental decisions constrained by limited budgets (e.g. Hanson et al., 2019, Castonguay et al. 2023, Timms and Holden, 2024), and is increasingly used to inform several aspects of fisheries management. For example, several studies have analysed costs and benefits of implementing broad management strategies (Changeux et al., 2001; Dowling et al., 2013; Mangin et al., 2018; McGowan et al., 2018, Erm et al 2023) and of collecting better and more frequent biological data (Dennis et al., 2015; Prellezo, 2017; Punt et al., 2002, Hutniczak et al, 2019, Bisack and Magnusson, 2014).

However, we are aware of only one study that examines both costs and benefits of modelling approaches. Dichmont et al. (2017) considered outcomes resulting from the application of single-species assessment models of different complexities in a mixed fishery. They found that the benefit of age-structured integrated models and associated data outweighed the cost of both implementing the more complex models and the cost of collecting the required data to parameterize them. Simple methods reduced benefits through: (1) degraded stock states and associated foregone future catch, valued at orders of magnitude larger than explicit assessment costs; or (2) the need for precautionary harvest rules to account for simple assessment uncertainty, leading to unrealised catch. Either way, model complexity benefits outstripped costs. Note, however, that Dichmont et al. (2017) did not separate the costs and benefits of collecting the data to parameterize the more complex models from the modelling costs, and they did not consider ecosystem models (e.g. Fig. 1 d-e). Furthermore, this type of cost-benefit analysis is strictly in monetary units, which may be inappropriate for quantifying many of the benefits of ecosystem modelling.

Rather than converting benefits to monetary units and simply subtracting or dividing benefits and costs, an alternative approach would be to consider the benefit of each model against its associated cost. The boundary (highest benefit) at each cost value forms a Pareto frontier (Castonguay et al 2023), which allows stakeholders to weigh trade-offs between costs and benefits. The shape of this frontier would differ depending on the objectives and models considered, especially for benefits not easily translated into monetary units, such as effects on at-risk biodiversity or human well-being. If costs and benefits can be subdivided into two categories – benefits easily translated into monetary units and benefits not easily translated to monetary units – the cost-benefit analysis can be visualized as a trade-off curve (pareto frontier) between non-monetary and monetary benefits. This allows managers and stakeholders to examine their own trade-offs between costs and benefits, regardless of units. There has been an increasing uptake of such multi-criteria decision analysis approaches to compare the benefits of fisheries management decisions (Bisack and Magnusson, 2014, Pascoe et al. 2017, Pascoe et al. 2023). However, they have yet to be used to explore the benefits and costs of different modelling frameworks.

Below we illustrate how such approaches can be used to compare the costs and benefits of model complexity for fisheries management by working through a hypothetical example using our relative cost data from Fig 2. In this example, consider an ecosystem with an unharvested predator, a targeted forage fish, a bycatch species, and several species which may be indirectly affected by fishing due to interactions. Now consider a fisheries modeler who informs a harvest rule recommendation for this fishery. They might, for example, use a production model for the target species (Fig 1a), and hypothetically their recommended harvest could deplete the fishery due to a lack of time lags and demographic structure.

Alternatively, they might perform an age-structured stock assessment (Fig 1b), leading to high values of sustainable harvest, but hypothetically still leaving the more sensitive bycatch species heavily depleted below B20, defined as 20% of biomass carrying capacity. B20 is the lower limit biomass reference point for Australian fisheries harvest policies. Pairing a production model for the bycatch species (Fig 1c) with either the age structured or single species production model for the harvest species might rescue the bycatch species above B20, but this does not account for indirect effects of fishing, potentially leaving other species vulnerable or leading to dynamics opposite to what results from nonlinear indirect effects (Table 1, last column, rows 2 and 4). For example, if the modeler developed an ecosystem model of intermediate complexity (Fig 1d) including the predator, they might have recognised that harvest needed to be reduced further to keep the predator above B20 (assuming for simplicity that the same reference levels are appropriate for each species, Table 1, last column, row 5). Lastly, in this hypothetical example, if the modeler developed an even more complex ecosystem model, they might have uncovered an unexpected, additional vulnerable species that could decline below B20 through nonlinear indirect links down the trophic chain. In the table below, we use the cost ratios between different modelling approaches (Fig 2) along with hypothetical long-term values of catch (Table 1, column 3) to enumerate our example. We then use this table to walk through a hypothetical cost-benefit analysis (Fig. 3).

**Table 1.** Costs and hypothetical benefits of fisheries models used for the cost-benefit analysis example in Fig. 3. Relative model costs are scaled in line with relative costs as in Fig 2.

| Model | Model cost ($100k) | Value of catch ($100k) | No. species conserved > B20 |
|---|---|---|---|
| Production | 0.3 | 15 | 0 |
| 2 sp. Production | 0.6 | 35 | 2 |
| Age structure | 1.0 | 50 | 1 |
| Age structure + bycatch | 1.3 | 48 | 2 |
| MICE | 3 | 45 | 3 |
| Complex Eco. Model | 15 | 40 | 4 |

With values for benefits and costs in hand, one can do a cost-benefit analysis. In the hypothetical case here, we recommend the trade-off analysis where one plots benefits minus costs in monetary units against the number of species rescued. In Fig 3, a manager only concerned about the monetary gains would select an age-structured assessment (left endpoint of the dashed pareto frontier in Fig. 3). However, one could save one and two more species, with little sacrifice to monetary gains by either adding a bycatch production model or developing a MICE, respectively (light grey points in the middle of the dashed pareto frontier in Fig. 3). At an added cost, the modeler could further identify a harvest rule that saves an additional species by conducting a more complex ecosystem model (white point on the pareto frontier in Fig. 3). In this hypothetical example, one and two species production models are dominated modelling strategies, that sit below the optimal pareto frontier. The single species production model leads to both low monetary benefits (due to low catch) and zero species saved. The two species production model, while saving two species, could be improved, by increasing monetary value (with no sacrifice to species saved, moving vertically in Fig. 3) or increasing the number of species above B20 (with no sacrifice to monetary value, moving to the right, and up, in Fig. 3). These two improved outcomes are achieved by adding age structure to the target species model, or developing a MICE, respectively. Whether to use an age-structured model, age-structure with bycatch, MICE, or a complex ecosystem model would depend on how the stakeholder values monetary gains versus species saved.

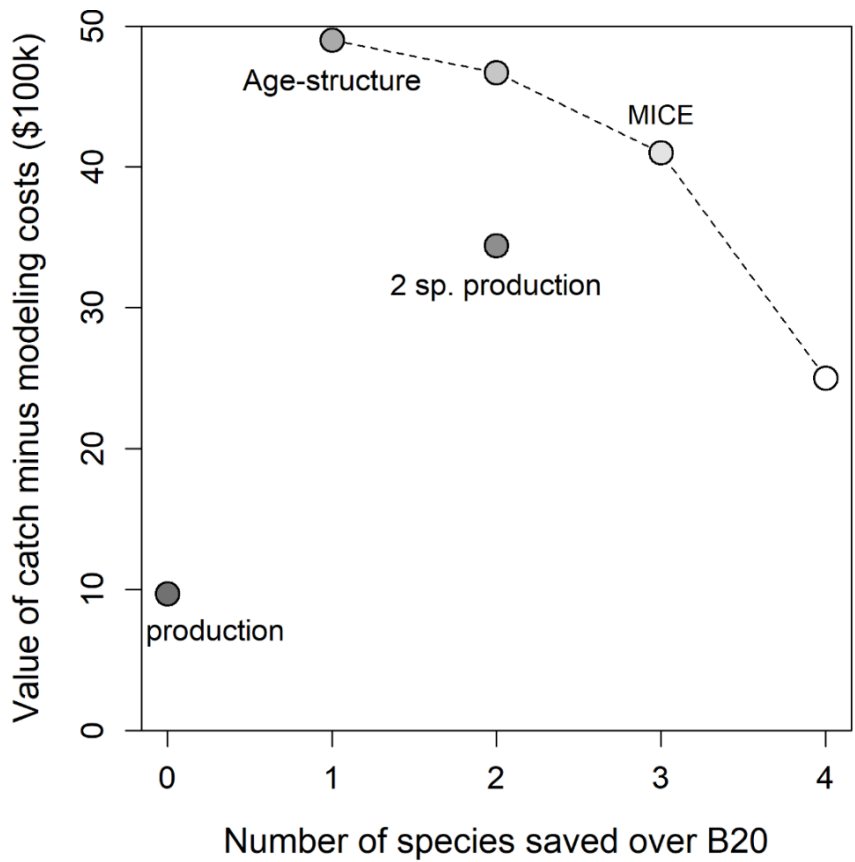

**Fig. 3.** Pareto trade-off frontier for the hypothetical fishery presented in Table 1. Dark circles represent the benefits of low-cost simple models, and the white point represents the benefits of developing a complex ecosystem model. The light grey circles on the dashed curve are models of medium levels of complexity (Age structure + bycatch, and MICE).

It is important to emphasize that fisheries scientists and managers may have a great deal of uncertainty in all quantities displayed in Table 1 and Fig. 3 for real world fisheries. However, managers can adjust values in the table and see how the points move around in Fig.3. Doing so, before models are ever built or run, may still be able to suggest which models are likely to be most cost-effective. Often dominated strategies (those not on the pareto frontier) that will perform poorly on all measures (such as the production and 2 species production model in Fig. 3) are robust to major deviations in the problem specification. Just how walking through the assumptions and logic behind a model is beneficial, independent of the model, the same is true for structured decision making. Walking through cost-benefit analysis logic can be valuable, even if benefits and costs are largely unknown.

**Conclusion**

Debates around best practices for incorporating model complexity are common (Karp et al., 2023). Recommendations include stepwise addition of model complexity, and only increasing complexity when justified (e.g. by the fit to the data), and to use simpler models if available data are noisy (Yearsley & Lettenmaier, 1987). The few fisheries studies that have tested model complexity against field data suggest that intermediate levels of complexity predict future data best (Fulton et al., 2003; Håkanson, 1995). But the focus has been largely around precision and accuracy, rather than management outcomes (Fulton et al., 2019).

Improved precision and accuracy resulting from appropriate model complexity (Håkanson, 1995) could improve management outcomes. For example, managers often make conservative decisions to protect a fish stock given high levels of uncertainty (Mildenberger et al. 2021). Reducing uncertainty through better data or models could lead to less precautionary management decisions that in turn increase the profitability of the fishery (Holzer 2017). However, it is important to note that improved accuracy and precision of predictions does not always lead to better management (Boettiger, 2022). Even when they do such improvements can come at a substantial cost.

We argued here that the value of model complexity should be examined relative to costs and have encouraged reporting of costs and benefits of modelling for fisheries management. Such transparency could allow agencies to better understand how modelling supports and addresses management needs and could also inform best practices in the industry. To help facilitate the use of cost benefit analysis in fisheries modelling we recommend the following steps:

**Step 1: Define objectives (benefits) and possible actions (management decisions).** The choice of model will depend on the questions of interest, objectives, and available policy options. Quantifiable metrics for benefits could include, economic metrics, such as revenue, or number of employed fishers. Metrics could also be environmental, for example, number of species conserved over a threshold target, or geometric or arithmetic mean of species abundances or biomasses (Erm et al 2023), or total carbon emissions (Castonguay et al 2023), or pollution averted. Other metrics could even focus on research benefits such as, new hypotheses generated for refinement and testing, total grant revenue awarded, number of citations, or new software developed. Some benefits such as cultural and social ones may be harder to quantify.

**Step 2: Identify available data, and a set of candidate models.** Some models may require more data to be reliably parameterized. However, even overly-complex or overfit models may be helpful for scenario analysis. It is important to reflect on whether more data is needed to achieve the objectives. We have largely focussed on the perspective of a need for modelling in the absence of control over what data are available. However, in an ideal scenario, candidate models could be considered in tandem with the decision to collect more data. In some cases, extensive data collection may be more important than model choice, as highly informative data can yield cost-effective simple harvest control rules.

**Step 3: Reflect on likely costs and feasibility of model implementations.** Choose one or more modelling frameworks to meet objectives most effectively relative to budget constraints. Ideally this selection would include a consideration of both costs and benefits using formal cost-benefit analysis. It is important to note that in many cases it would be costly and logistically unreasonable to build, run, and analyse every candidate model. There are methods to help conduct such analyses without having to run such models, e.g. via expert elicitation workshops (Pascoe et al 2017). However, if management agencies reported costs and benefits achieved from modelling, anyone could do a hypothetical cost benefit analyses quickly and cheaply using costs and benefits reported in the literature.

**Step 4: Report costs and benefits.** Costs and benefits (in terms of the stated objectives) of each approach attempted (even if only one approach is attempted). Predicted benefits could be generated by projecting models forward under new recommended decisions and comparing benefit metrics (see step 1) to outputs from the models run under business as usual or model-free scenarios.

Step 4 is perhaps the most important. Often fisheries managers and agencies do not know precise modelling costs going into a project, especially if they are deciding among a suite of candidate models. Similarly, benefits of different modelling approaches are often highly uncertain. In fact, the reason why our example in Fig. 3 is hypothetical is because we are aware of no study that has ever reported simultaneously, both the costs and benefits of using a model. Many agencies would benefit from the known capacity of models of a specific type to deliver reliable cost-effective results for the question being asked. Fisheries science focuses on achieving more for less – higher yields, better outcomes for ecosystems, and increased sustainability. Using cost-benefit analysis to examine the value of complex models will further the pursuit of efficiency into fisheries modelling practice.

**Acknowledgements**

Funding provided by the Australian Research Council grant DE190101416

**Data Availability Statement**

Raw cost data are sensitive and contained in private legal contracts between organizations and their clients. However, relative cost ranges are directly displayed in Fig 2.

**Conflict of Interest Statement**

The authors have no conflicts of interest.